\documentclass[12pt,preprint]{aastex}

\def\etal       {et al.~}

\def\threesig   {$3\sigma$}

\def\AIPS      {{\small AIPS}}
\def\IMAGR     {{\small IMAGR}}
\def\CLEAN     {{\small CLEAN}}
\def\uvplane   {{\it(u,v)}-plane}
\def\uvdist    {{\it(u,v)}-distance}

\def\hho       {H$_2$O}
\def\Htwo      {H$_2$}
\def\sioone    {SiO v$=1$, J=1--0}

\def\rstar     {\ifmmode {R_*} \else {$R_*$} \fi}
\def\oceti     {$o$~Ceti}
\def\rleo      {R~Leo}
\def\whya      {W~Hya}

\def\Hminus    {H$^-$ free-free}

\def\mjyb   {\ifmmode {{\rm mJy~beam}^{-1}} \else{mJy~beam$^{-1}$}\fi}
\def\jyb    {\ifmmode {{\rm Jy~beam}^{-1}} \else{Jy~beam$^{-1}$}\fi}

\def\Vorb    {\ifmmode {V_{orb}} \else ${V_{orb}}$ \fi}
\def\Vdot    {\ifmmode {\.V} \else ${\.V}$ \fi}

\def\arcm{\ifmmode {' }\else $' $\fi}
\def\arcs{\ifmmode {'' }\else $'' $\fi}
\def\arcmper{\ifmmode \rlap.{'} \else $\rlap{.}' $\fi}
\def\arcsper{\ifmmode \rlap.{''} \else $\rlap{.}'' $\fi}
\def\porm   {\ifmmode\pm\else$\pm$\fi}
\def\kms    {\ifmmode{{\rm ~km~s}^{-1}}\else{~km~s$^{-1}$}\fi}
\def\peryr  {\ifmmode{{\rm ~yr}^{-1}}\else{~yr$^{-1}$}\fi}
\def\gmpercc   {\ifmmode{{\rm gm~cm}^{-3}}\else{gm~cm$^{-3}$}\fi}
\def\percmcube   {\ifmmode{{\rm ~cm}^{-3}}\else{~cm$^{-3}$}\fi}
\def\percmsquare {\ifmmode{{\rm ~cm}^{-2}}\else{~cm$^{-2}$}\fi}
\def\masy   {\ifmmode{{\rm mas~y}^{-1}}\else{mas~y$^{-1}$}\fi}
\def\micron {\ifmmode{\mu{\rm m}}\else{$\mu$m}\fi}
\def\a      {\ifmmode {\rlap.}^{''}\! \else ${\rlap.}^{''}\!$\fi}
\def\p      {\phantom{0}}

\newbox\grsign \setbox\grsign=\hbox{$>$} \newdimen\grdimen \grdimen=\ht\grsign
\newbox\laxbox \newbox\gaxbox
\setbox\gaxbox=\hbox{\raise.5ex\hbox{$>$}\llap
     {\lower.5ex\hbox{$\sim$}}}\ht1=\grdimen\dp1=0pt
\setbox\laxbox=\hbox{\raise.5ex\hbox{$<$}\llap
     {\lower.5ex\hbox{$\sim$}}}\ht2=\grdimen\dp2=0pt
\def\gax{\mathrel{\copy\gaxbox}}
\def\lax{\mathrel{\copy\laxbox}}

\slugcomment{2007 July 20}

\shorttitle{Imaging Mira Radio Photospheres}

\shortauthors{Reid \& Menten}

\begin{document}

\title{Imaging the Radio Photospheres of Mira Variables}

\author{M. J. Reid}
\affil{Harvard--Smithsonian Center for Astrophysics,
    60 Garden Street, Cambridge, MA 02138}
\email{reid@cfa.harvard.edu}

\author{K. M. Menten}
\affil{Max-Planck-Institut f\"ur Radioastronomie,
       Auf dem H\"ugel 69, D-53121 Bonn, Germany}
\email{kmenten@mpifr-bonn.mpg.de}

\begin{abstract}
      We have used the VLA at 43 GHz to image the radio continuum
emission from \oceti, \rleo, and \whya\ and to precisely locate their SiO
maser emission with respect to the star.  The radio continuum emission
region for all three stars has a diameter close to 5.6~AU.  These diameters
are similar to those measured at infrared wavelengths in bands
containing strong molecular opacity and about twice those measured
in line-free regions of the infrared spectrum.  Thus, the radio photosphere 
and the infrared molecular layer appear to be coextensive.
The 43~GHz continuum emission is consistent with temperatures near 1600~K 
and opacity from \Hminus\ interactions.  While the continuum image of
\oceti\ appears nearly circular, both \rleo\ and \whya\ display 
significant elongations.  The SiO masers for all three stars show 
partial rings with diameters close to 8~AU.  
\end{abstract}

\keywords{Stars: Long-Period Variables, Circumstellar Matter, 
Atmospheres -- Radio Continuum: Stars -- Masers}

\section{Introduction}

A typical oxygen-rich Mira variable has a radius, \rstar, defined by
line-free regions of the spectrum, of about 1--2 AU.  Surrounding
the star is an extensive ``molecular layer,'' which extends
to nearly 2\rstar\ and can
be opaque across most of the visible and infrared spectrum.   
The time-variable formation of metallic oxides in the molecular
photosphere can explain the thousand-fold (8 mag) variations in
the optical light curves of some Miras \citep{RG02}.
Also at about 2\rstar\ one detects a
``radio photosphere,'' surrounded by an SiO maser emitting region.
Beyond the SiO region dust grains form, radiation pressure
accelerates the dust, transferring
outward momentum to the gas and leading to significant mass loss.

\citet{RM97a}
published measurements made with the Very Large Array
(VLA) of a sample of Mira-like variables
(\oceti, \rleo, \whya, R~Aql, $\chi$~Cyg, and R~Cas).
Their results revealed that the these stars exhibit a cm-wavelength 
spectral index of $\approx 1.9$ and nearly constant radio light curves 
(variation $\lax\pm15$\% about the mean).
Also, they imaged the radio continuum of \whya\ at 22 GHz and found
the star precisely at the center of a ring of
\hho\ masers with a radius of 0\a15 \citep{RM90}.

The detection of a ``radio photosphere" near $2R_*$ provides a new
observational tool to increase our understanding of the region
between the star and the dust formation zone.
For the physical conditions expected in the radio photosphere,
free electrons, obtained mostly from the ionization of potassium
and sodium, provide the dominant opacity through free-free
interactions with neutral H and \Htwo.  \citet{RM97a} proposed
a simple physical model which can explain the radio frequency observations.
At cm-wavelengths, unity optical depth is achieved at a radius of
about 3 to 4~AU, where the density and temperature
are $\sim10^{12}$~\percmcube\ and $\approx1600$~K, respectively.
This produces a flux density which is roughly twice that expected for
black-body emission from the underlying star, as is observed.

In this paper, we report observations at 7~mm wavelength (43 GHz) 
in which we simultaneously imaged the radio photospheres and SiO maser emission
from two Miras (\oceti\ and \rleo) and one semi-regular variable (\whya).  
These observations and supporting models constrain the
physical and dynamical conditions between $2$ and $3~\rstar$

\section{Observations \& Data Analysis}

The observations were made with the NRAO
\footnote{The National Radio Astronomy Observatory (NRAO) is operated 
by Associated Universities, Inc., under a cooperative
agreement with the National Science Foundation.}
Very Large Array (VLA) under program AR446.
Our goal was to measure directly the angular size of radio photospheres
with the highest angular resolution afforded by the VLA.
Our synthesized beam at 43 GHz in the A-configuration was
$\approx 0\arcsper04$ full-width at half-maximum (FWHM),
which well resolves a typical Mira to a distance of $\approx200$~pc.
Since radio frequency ``seeing" is limited primarily by fluctuations of
water vapor in the atmosphere and is rarely better than $0\arcsper1$,
we used a calibration scheme developed by us \citep{RM90,RM97a}
in which circumstellar masers provide a phase reference for the 
continuum data.  This allows diffraction limited imaging to be achieved.

Between 2000 October 25 and November 6, we observed three stars:
\oceti, \rleo, and \whya.  Optical stellar phases for the three stars
were approximately 0.05, 0.55, and 0.25, based on examination of
AAVSO visual magnitudes for several cycles before and after the observations.

The observations to measure stellar sizes 
were conducted in 4-IF continuum mode.  The A-C IF pair 
(50 MHz bandwidth in right and left circular polarization) was centered 
at 43164.9 MHz, which is 43~MHz above the rest frequency
of 43122.08 MHz of the \sioone\ line.
The B-D IF pair used a 1.563~MHz bandwidth centered on the SiO
line at local standard-of-rest (LSR) velocities of
47.0~\kms\ for \oceti, 1.0~\kms\ for \rleo, and 42.0~\kms\ for \whya.

The interferometer data were calibrated using the Astronomical Image 
Processing System (\AIPS).
Absolute flux density calibration was obtained from an observation of 
3C286, assuming a flux density extrapolated from the formula
given by \citet{B77}.
Following the procedures described in detail in \citet{RM97a},
we used roughly hourly observations of a strong continuum source to
update pointing, to remove the electronic phase differences among the bands
and to monitor the slowly-varying amplitude response of the antennas.
The narrow-band data, containing very intense SiO maser emission, 
were phase-only self-calibrated on each 10~sec integration.  
The resulting antenna--dependent
phase corrections were then applied to the broad-band data
containing the continuum emission from the star.
Since the masers are observed simultaneously with the continuum data
and come from essentially the same position on the sky,
this calibration removes all significant atmospheric phase-fluctuations.  

\begin{figure}
\epsscale{0.65}
\plotone{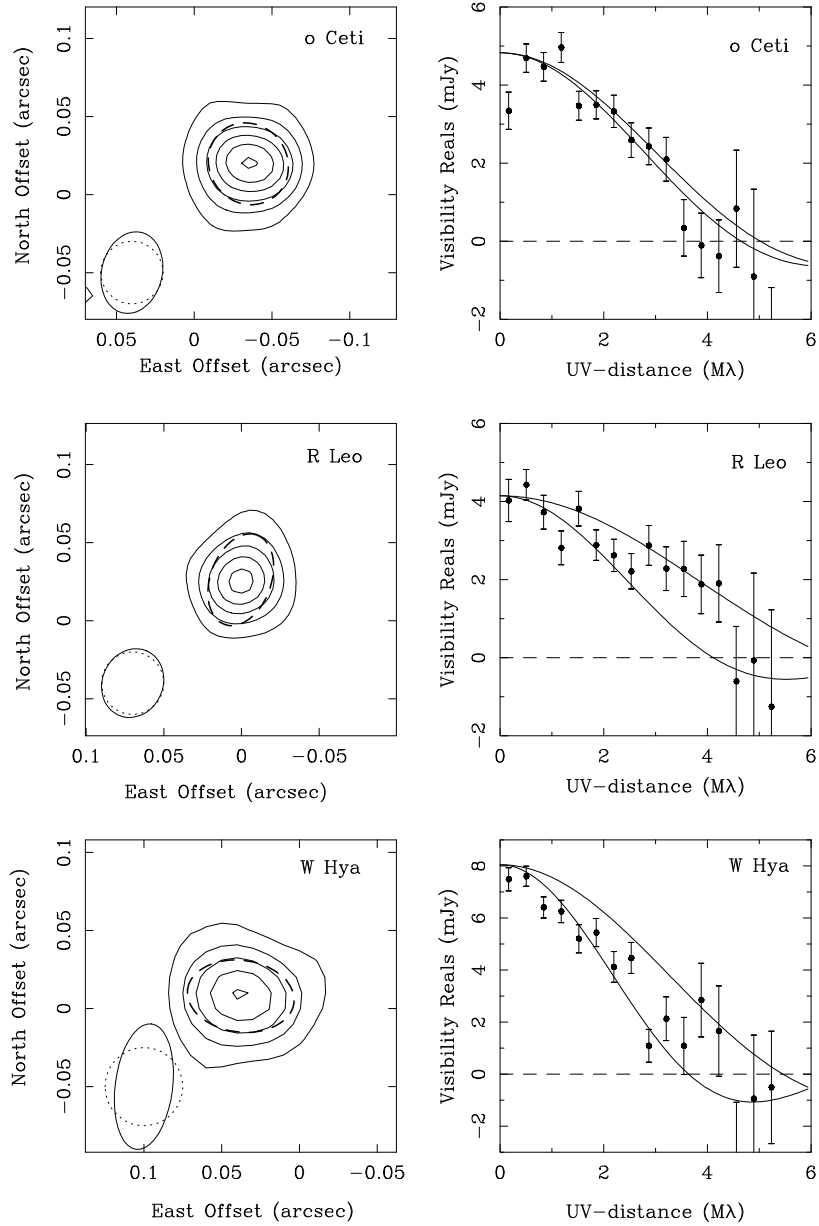}
\caption{Images shown as contours ({\it left panels}) and 
visibility vs. interferometer baseline length ({\it right panels}) 
of the 43 GHz continuum emission from \oceti\ ({\it top panels}), 
\rleo\ ({\it middle panels}) and \whya\ ({\it bottom panels}).
Contour levels are integer multiples of 0.5 mJy for \oceti\
and \rleo\ and multiples of 1.0 mJy for \whya.
In the image panels, the interferometer FWHM ``dirty'' beams are shown as 
solid ellipses and the round restoring beams are shown as dotted 
circles in the lower left.  
The dashed ellipses superposed on the images are the sizes of
uniformly bright elliptical disks that best fit the data.
In the visibility plots, the two model curves correspond to 
expected visibilities for circular disks with diameters 
corresponding to the major and minor axes listed in 
Table~\ref{table:fit_parameters}.
   \label{fig:star_visibilities}
        }
\end{figure}

Once the data were ``cross-self-calibrated,'' both the narrow-band maser 
and broad-band continuum signals were imaged with the \AIPS\ task \IMAGR.  
Map resolutions and noise levels are given in 
Table~\ref{table:map_parameters}.
In order to simplify map analysis, \CLEAN\ maps used a round restoring 
beam equal approximately to the geometric mean of the dirty beam's 
major and minor axes.  These maps are shown in left column of panels in 
Fig.~\ref{fig:star_visibilities}.  

In addition to images, we also analyzed the interferometer visibility 
data in the following manner.  
Firstly, we measured the position of peak continuum emission by fitting 
an elliptical Gaussian brightness distribution to the map.
Next, we ``shifted'' the data to place the continuum
peak at the interferometer phase-center.   After time averaging
the visibilities for 1~minute, we fitted the data in the \uvplane\ 
with a uniform-brightness, elliptical-disk model.  
This fitting was done using MIRIAD analysis software.
Parameters of these fits are given in Table~\ref{table:fit_parameters}. 

Displaying the 2-dimensional \uvplane\ data and model fits for
weak sources is difficult.   In order to obtain sufficient 
signal-to-noise ratio, the data need to be vector averaged 
over large portions of the irregularly sampled \uvplane.
This precludes, for example, plotting visibility amplitude
versus projected baseline length along a narrow strip at
given position angle in the \uvplane.   
We have chosen to bin the data and plot
the real part of the visibility versus baseline length.
These plots appear in the right-hand panels of 
Fig.~\ref{fig:star_visibilities}.
For an elliptical source, this mixes data with differing
visibilities, owing to the greater resolution of a source
for a given baseline length if it is aligned with the 
source major axis than the minor axis.  
In order to indicate the magnitude of this effect, we have  
plotted the theoretical visibilities for circular sources
with diameters equal to the major and minor axes of the
fitted elliptical model.

We also observed the SiO maser emission at high spectral resolution 
with several scans in spectral-line mode interspersed among the
dual-band continuum observations.   We covered the entire SiO line
emission with a bandwidth of 3.125 MHz and used 64 spectral channels, 
which provided a velocity resolution of 0.34 \kms.  The line data
were self-calibrated by choosing a channel with strong emission
as a reference, and the resulting phase and amplitude corrections 
were applied to the other channels.  Because the maser emission was
very strong ($\sim10^3$ Jy), we did not require bandpass calibration.  
From the line data we produced a spectral-line data cube, which we
restored with a 20~mas FWHM Gaussian beam, taking advantage of the high
signal-to-noise ratio and nearly point-like emission in each spectral
channel to ``super-resolve'' by a factor of about two.

\begin{figure}
\epsscale{0.40}
\plotone{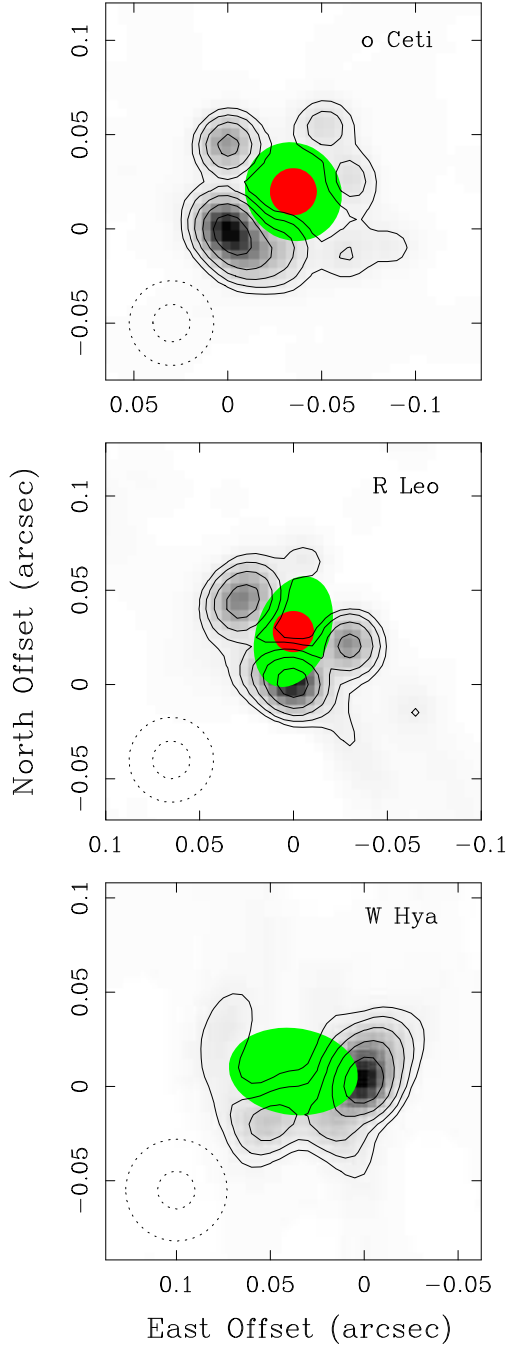}
\caption{SiO maser emission ({\it contours and gray scale)} 
superposed on the fitted uniformly bright elliptical disk models 
of the radio continuum emission ({\it green filled ellipses}) from 
\oceti\ ({\it top panel}), \rleo\ ({\it middle panel}) and \whya\ ({\it bottom panel}).
SiO contour levels are 1, 2, 4, 8, 16, and 32 times 100, 40,
and 200 \jyb\ for the three stars, respectively.
The relative positions of the SiO masers and the radio continuum
have been measured with accuracies of about 2~mas.  The restoring
beams for the radio continuum (larger dotted circle) and SiO 
maser data (smaller dotted circle) are indicated in the lower 
left corner of each panel.
The true stellar sizes ({\it red filled circles}) are from
circular, uniformly bright, disk models of IR interferometer data 
\citep{P04}.
   \label{fig:star_maser_overlays_grey}
        }
\end{figure}

The continuum emission (in the 50 MHz A-C IF pair) and maser emission
(in the 1.563 MHz B-D IF pair), which were observed simultaneiously, 
are automatically aligned by the calibration procedures to an accuracy
of better than 1~mas.  In order to align the maser emission from the
spectral-line data, not observed simultaneously, with the continuum,
we performed the following steps: 
1) we produced a pseudo-continuum map from 
the line data, using spectral channels covering the velocities that were 
within the 1.563 MHz passband of the dual-band continuum setup; 
2) we overlaid the pseduo-continuum map and on the map from the 1.563 MHz 
passband; and 
3) we translated the pseduo-continuum map to achieve the best alignment
by eye.  Since the maps were nearly identical and contained unresolved 
spots with high signal-to-noise ratios, alignment of the continuum and 
maser emission to better than $\approx2$~mas (or 10\% of the restoring 
beam) was achieved.  
Based on this cross-registration, we show velocity-integrated SiO maser 
emission superposed on the continuum emission of the stars in 
Fig.~\ref{fig:star_maser_overlays_grey}.
In order to avoid introducing excess noise in the velocity-integrated 
SiO maser maps, we set the flux density of all pixels with less
than 1~\jyb\ to zero before summing the emission of the
individual spectral channels.

\section {Radio Photospheric Characteristics}

\subsection{Diameters}

We clearly resolve the radio photospheres of all three stars.
For \oceti, we find a nearly circular image with a uniform-disk
diameter of about 52~mas (see Table~\ref{table:fit_parameters}).  
At a distance of 110~pc, based on the Period-Luminosity (PL) 
relation of \citet{F89} and the photometry of \citet{H95}, 
this corresponds to a radio diameter of 5.7~AU.  
Throughout this paper, we adopt the PL-based distances,
which appear to be more accurate than {\it Hipparcos} parallaxes 
for these large, variable and irregularly bright stars \citep{RM97a}.
We measure a mean diameter of 50~mas for \rleo, 
corresponding to 5.5~AU at a distance of 110~pc.  
\whya\ displays a mean angular diameter 
of 58~mas, which at 95~pc distance implies a physical diameter 
of 5.5~AU.  Thus, the radio-frequency mean diameters of all three
stars are remarkably similar, averaging close to 5.6 AU.

The diameters of the radio photospheres are similar to 
those measured in the visible and infrared in narrow bands
that contain strong molecular opacity.  For example, in
a 10-\micron-wide band centered at 710~\micron\ containing
strong TiO absorption, \citet{H95} find (uniform-disk) angular 
diameters of 53, 45 and 66~mas for \oceti, \rleo\ and \whya, 
respectively.  Similarly, \citet{P04} infer a ``molecular layer''
diameter of 52 and 50~mas for \oceti\ and \rleo, respectively.
Generally it is difficult to ``see'' through the molecular
photosphere and down to \rstar.  Using narrow bands placed
in spectral regions with minimal molecular opacity, \citet{P04} 
find stellar diameters of 26~mas for \oceti\ and 22~mas for \rleo.  
Model-fit Rosseland mean diameters from observations using broader 
bands yield diameters that are $\approx20$\% larger \citep{W04,F05}.  
Thus, the radio and molecular layers occur at radii of
$\approx2\rstar$, adopting the narrow-band stellar diameters.

\subsection {Mira Binary}

Mira A is \oceti, an M-type giant, and Mira B is possibly a low-mass
main-sequence star \citep{I06} or a white dwarf.  \cite{MK06} detected 
both stars in the Mira binary with the VLA.  They found the 43~GHz flux
density of Mira B is weaker than that of \oceti\ by a factor
of $\approx3$.  We well resolve the binary, which has a projected
separation of $\approx500$~mas, and fail to detect Mira B.
Our \threesig\ limits are 0.4 and 0.7~\mjyb\ for untapered and
tapered maps with (geometric mean) interferometer beams of 50 and 250~mas, 
respectively.
Were the radio emission from Mira B unresolved and at the
$\approx0.9$~mJy level seen by Matthews \& Karovska, we would have
detected it.   Since the Matthews \& Karovska B-configuration
observations at 43~GHz had a beam size comparable to our tapered
map, it appears that angular resolution alone cannot explain 
our failure to detect Mira B and that it must be time variable.
Since we well resolve ($\approx10$ beams) the binary and Mira B
is not detected, our \oceti\ results should not be significantly
affected by the secondary.

\subsection{Asymmetries}
  
While \oceti's image is circular to within $\pm4$\%, we  
detect non-circular radio photospheres for \rleo\ and \whya.
The differences between the major and minor axes are significant at 
about the $2\sigma$ level.  
We are reasonably confident that the non-circular structures 
do not come from instrumental limitations for several reasons:
1) for \rleo\ the intrinsic interferometer beam is nearly round
and for \whya\ the image is elongated nearly perpendicular to the
intrinsic interferometer beam;
2) analysis of the data both in the image and \uvplane\ yield similar
results; and
3) the plots of visibility versus \uvdist\ for the elliptically shaped
stars show scatter larger than expected for a circular source, but
consistent with an elliptical source.

\subsection{Physical Conditions}

Both \oceti\ and \rleo\ have (uniform-disk) brightness temperatures
near 1650~K.  These temperatures are close to those expected for
the radio photospheres \citep{RM97a}.
We measure a somewhat higher brightness temperature of $2380\pm550$~K 
for \whya.  
Since the collisional ionization of metals, and hence the \Hminus\ opacity, 
is strongly temperature dependent below 2500~K, the higher measured
brightness temperature of \whya\ might suggest that the total density in 
the \whya\ photosphere is lower than in \oceti\ or \rleo.
However, the uncertainties in brightness temperature are fairly large,
owing predominantly to the squared-dependence on the uncertainty in the 
size measurement, and all three stars could have similar temperatures.
Our temperature measurements are consistent, within uncertainties, 
with our earlier measurement of brightness temperature of \whya\ of 
$1570\pm570$~K at a lower frequency of 22 GHz \citep{RM97a}.   

The brightness temperatures of the radio photospheres of \oceti\ and 
\rleo\ are also similar to those inferred by \citet{P04} of about 2000 
and 1600~K for the molecular layers of these stars.  
Spectroscopic observations of \hho\ in the infrared 
also indicate temperatures and densities close to those we previously 
measured and modeled.  
\citet{M02} model their 2.5 to 4.0~\micron\ ISO telescope data
with a hot (2000~K) layer close to the stellar photosphere 
(between 1 and 2\rstar) and a cool (1400 to 1000~K) layer 
between 2.5 and \hbox{4\rstar.}  The total hydrogen densities in these 
layers are estimated to be $\sim10^{12-13}$~\percmcube.  
These layers roughly straddle the radio photosphere where
we have previously found temperatures of about 1600~K and
hydrogen densities of $\sim10^{12}$~\percmcube\  
\citep{RM97a}. Since both the radii, temperatures and densities 
measured in the radio and infrared agree, this provides added confidence 
in both measurements.  

\section{SiO Masers}

The SiO masers in \oceti, \rleo\ and \whya\ form partial rings
with radii of about 40, 35 and 41~mas, respectively
(Fig.~\ref{fig:star_maser_overlays_grey}). 
These radii were measured by visually overlaying circles of varying
radii centered on the stellar continuum position and are probably 
accurate to $\pm10$\%, limited mostly by the lack of complete maser rings.
These partial rings are consistent with, but perhaps slightly larger than, 
the Very Long Baseline Array (VLBA) results by \citet{C04}, 
who find radii at different epochs ranging from 30--38, 26--30, and 42~mas 
for  \oceti, \rleo\ and \whya, respectively, for the \sioone\ transition .
The differences in ring radii between our VLA images, which are sensitive 
to lower brightness emission, and the VLBA results, which have
higher angular resolution and use a different approach to define
the ring radii, are small.  We note that \citet{DK03}, in their
extensive VLBA monitoring of TX Cam, find lower brightness SiO
masers radially outside of the brightest masers.  

\begin{figure}
\epsscale{0.80}
\plotone{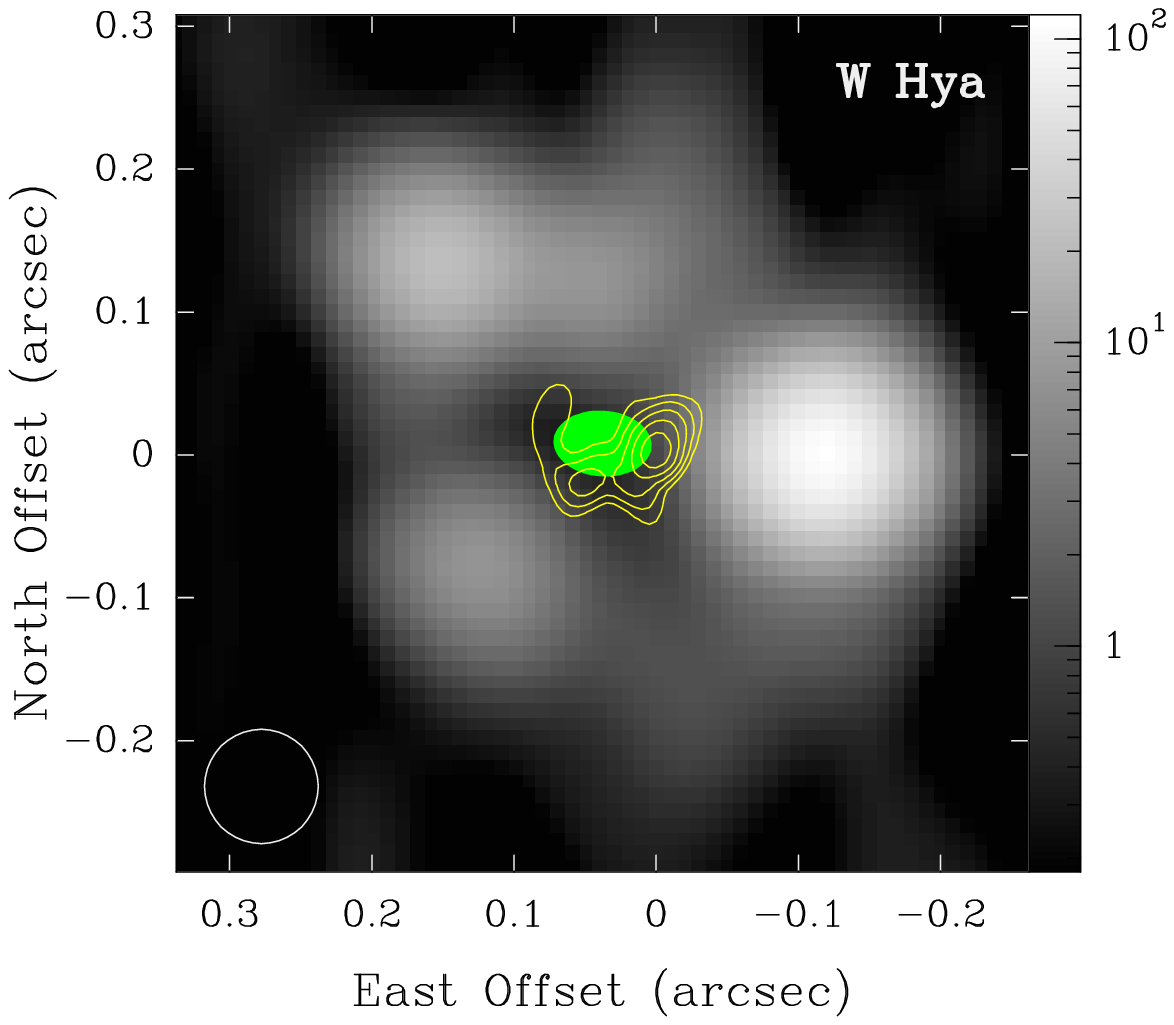}
\caption{\whya\ \hho\ maser emission ({\it gray scale)} from \citet{RM90},
superposed on the SiO maser emission ({\it yellow contours)} and the
radio continuum emission (elliptical disk model)
({\it green filled ellipse}) from Fig.~\ref{fig:star_maser_overlays_grey}.   
The \hho\ velocity-integrated flux scale in \jyb\  
is shown along the right side of the figure.
The relative positions of the \hho\ masers and the radio continuum
is accurate to about 10~mas.  The restoring
beam for the \hho\ emission is indicated in the lower left corner.
   \label{fig:whya_h2o}
        }
\end{figure}

When we convert our ring diameters from angular to linear units,
we find radii of 4.4, 3.8 and 3.9~AU, respectively.   
Thus, these three stars show very similar SiO masing radii 
close to 4~AU.
Previous observations have also detected partial rings of SiO masers
in Mira-like stars \citep{M94,G95,B00,DK03} and have usually 
{\it assumed} that the stars were near the center of the rings.  
However, for the first time, we are able to demonstrate that the 
maser rings are precisely centered on the stars. 
These results nicely complement our previous result that the
\hho\ masers in \whya\ also exhibit a partial ring and are
precisely centered on the stellar continuum emission \citep{RM90}.  
We plot the integrated \hho\ maser emission from \whya\ in 
Fig.~\ref{fig:whya_h2o}, superposed on its SiO 
maser and radio continuum emission.
The \hho\ maser ring had a radius of $\approx150$~mas, which is
about 3.7-times that of the SiO masers. 

The simplest interpretation for the SiO masers is that they form
in a spherical shell near radii of about 3\rstar.  If one adopts
the somewhat smaller SiO radii from VLBA maps, which sample the higher
brightness masers, and the larger stellar radii, from Rosseland
mean opacities, the maser shells would have radii closer to 2\rstar.  
The inner boundary of the SiO masers is determined either by
densities $\gax10^{10}$~\percmcube, leading to thermalization of the
level populations \citep{LE92}, or by high radio-continuum
opacity from \Hminus\ interactions \citep{RM97a}.
The outer boundary occurs when SiO densities become too low for
strong maser amplification.  Note that in addition to the rapid
decrease in total density with increasing radius in the extended
atmosphere, the SiO density can fall even faster
owing to silicate grain formation.

Proper motions of SiO masers in TX~Cam by \citet{DK03} indicate 
complex dynamics.  
Expansion at a speed of $\approx7~(D/390~{\rm pc})$~\kms, where
$D$ is the star's distance, is the dominant motion.
However, some SiO maser spots are also observed to move inwards
at a few \kms, suggesting a possible large-scale convective structure.
\citet{DK03} adopt a distance based on the period-luminosity and
photometry compiled by \citet{OWM01}.  An alternative distance estimate
could come from the radius of the SiO maser ring, since we find
all three of our stars have a ring radius near 4 AU.  Adopting this
ring radius for TX Cam, the measured mean angular radius of 16~mas
gives a distance of 250~pc.   At this distance the dominant SiO
expansion speed would only be 4.5 \kms.
In any event, there are no indications of pervasive, fast-moving 
($\gax10$~\kms) shocks in the SiO masing region.

\section{Discussion}

A simple, consistent picture of the inner envelopes of oxygen-rich 
Mira variables is emerging from radio and infrared observations.
Radio observations of stars like
\oceti, \rleo\ and \whya\ provide data which
constrains the kinematics and physical conditions between
$2$ and $3\rstar$.  Radio continuum emission
comes predominantly from \Hminus\ interactions.
An optical depth of unity is observed at $\approx2\rstar$
and requires a temperature and density of $\approx 1600$~K and
$\sim10^{12}$~\percmcube, respectively \citep{RM97a}.  

The radio photosphere corresponds very closely to the molecular layer,
identified at infrared wavelengths.   The molecular layer can
be directly observed in lines of \hho\ and CO \citep{H95,M02,P04}.
As the star cools cyclically,
this region can become nearly opaque at visible wavelengths,
owing to metallic oxide formation.  This leads to dramatic
changes in visible light as the temperature in the molecular
photosphere at $\approx2\rstar$ can be low enough for visible 
light to ``fall off'' the Wien side of the black body spectrum \citep{RG02}.

SiO maser emission surrounds the radio/molecular layer and   
appears as partial rings with radii of $2$ to $3\rstar$.  These rings
sometimes display significant ellipticity as, for example, shown
in TX Cam by \citet{DK03}.  
SiO masers trace both outwardly and inwardly moving material, possibly indicative
of large scale convection.  Temperatures and densities inferred
from SiO masers are $\approx 1000$~K and $\lax10^{10}$~\percmcube.
SiO maser proper motions suggest that outside of $3~\rstar$
shocks are mostly damped and propagate at $\lax5-7$~\kms.
This complements a 
similar conclusion, that strong, fast shocks are not found in the 
underlying radio photosphere, based on the observed nearly constant 
radio light curves \citep{RM97a,RM97b}.
Thus, while radio data indicate a dynamic region between $2$ and $3~\rstar$, 
there are no signs of strong shocks heating large portions of the atmosphere 
to $\sim10,000$~K near $2\rstar$ as in some models of these stars
\citep{W00}.

The ``radio photospheres'' of \oceti\ and \whya\ appear somewhat 
elongated.  The non-circular structure we detect in the radio 
photospheres of two AGB stars may be of significance in understanding 
the highly elongated structures seen in planetary nebulae,
following a post-AGB phase.  
Clearly second epoch images of the radio photospheres,
after several pulsation cycles, would be very interesting.  
If the non-circular structures change significantly, either in
shape or orientation, this would suggest a (non-radial) pulsation
origin.  Alternatively, should the structures be stable over time,
this would suggest a possible rotational or magnetic field origin,
which may play a role in creating asymmetric structures in 
planetary nebulae.

\acknowledgments
We thank L. Matthews for comments on an early version of the manuscript.
We acknowledge with thanks the variable star observations from the AAVSO 
International Database contributed by observers worldwide and used in this 
research.

\clearpage

\begin{deluxetable}{lccccc}
\tabletypesize{\scriptsize}
\tablecaption   {Interferometer Parameters \label{table:map_parameters}
                }
\tablehead{ \colhead{Star} &\colhead{$\theta_B(maj)$} &\colhead{$\theta_B(min)$} 
           &\colhead{P.A.} &\colhead{Continuum RMS Noise} &\colhead{Line RMS Noise}
         \\ &\colhead{(mas)} &\colhead{(mas)} &\colhead{(deg.)} 
            &\colhead{(mJy)} &\colhead{(mJy)}   
          }
\tablecolumns{6}
\startdata
\\
\oceti~~......... &53  &39 &$-13$  &0.13 &10 \\ 
\rleo~~.........  &61  &39 &$-20$  &0.14 &15 \\ 
\whya~~.........  &81  &37 &$-\p7$ &0.30 &20 \\ 
\enddata
\tablecomments{Beam major, $\theta_B(maj)$, and minor, $\theta_B(min)$, 
axes and East-of-North position angles, P.A., 
are from the best fit elliptical Gaussians 
to the ``dirty beams.''  The maps displayed in 
Fig.~\ref{fig:star_visibilities} and Fig.~\ref{fig:star_maser_overlays_grey}
used round restoring beams as discussed in the text. 
              }
\end{deluxetable}

\begin{deluxetable}{lcccccccc}
\tabletypesize{\scriptsize}
\tablecaption   {Measured Stellar Parameters \label{table:fit_parameters}
                }
\tablehead{ \colhead{Star} &\colhead{Distance} &\colhead{Opt. Phase} 
           &\colhead{$\theta_{maj}$} &\colhead{$\theta_{min}$} 
           &\colhead{P.A.} &\colhead{$S_\nu$} &\colhead{$T_b$} 
           &\colhead{SiO Radius} 
         \\ &\colhead{(pc)} & &\colhead{(mas)} &\colhead{(mas)} 
            &\colhead{(deg.)} 
            &\colhead{(mJy)} &\colhead{(K)} &\colhead{(mas)}
          }
\tablecolumns{9}
\startdata
\\
\oceti~~.........&$110\pm9$ &0.05 &$54\pm\p5$  &$50\pm5$ &$+39\pm50$  
          &$4.8\pm0.2$  &$1680\pm250$ &$40\pm4$ \\ 
\rleo~~.........&$110\pm9$  &0.55 &$61\pm10$  &$39\pm6$ &$-20\pm12$  
          &$4.1\pm0.2$  &$1630\pm410$ &$35\pm4$ \\
\whya~~.........&$\p95\pm8$ &0.25 &$69\pm10$  &$46\pm7$ &$+83\pm18$  
          &$8.0\pm0.4$  &$2380\pm550$ &$41\pm6$ \\
\enddata
\tablecomments{Distances are from \citet{H95} based on the PL 
relation of \citet{F89}. 
Stellar major, $\theta_{maj}$, and minor, $\theta_{min}$, 
diameters and East-of-North position angles, P.A., are from best-fit, 
uniform-brightness ($T_b$), elliptical-disk models.
Flux density, $S_\nu$, uncertainties do not include
a roughly $\pm10$\% uncertainty in the absolute flux density scale.
SiO maser shell radii are estimated from the brightness peaks in 
Fig.~\ref{fig:star_maser_overlays_grey}.  
              }
\end{deluxetable}

\clearpage

\clearpage

\clearpage

\end{document}